\documentclass[a4paper]{jpconf}

\usepackage{amssymb}
\usepackage{amsmath}
\usepackage{graphicx}
\usepackage{wrapfig}

\begin{document}
\title{Optimal driving protocols for nano-sized devices and their dependence on couplings to reservoirs}

\author{Mario Einax}

\address{Fachbereich Physik, Universit\"at Osnabr\"uck, Barbarastra\ss e 7, 49076 Osnabr\"uck, Germany}

\ead{meinax@uos.de}

\begin{abstract}
The development of efficient artificial nanodevices poses challenges which are of fundamental and technological nature. Recent progress has been
made in the context of finite-time thermodynamics. A central question in finite-time thermodynamics is to identify the optimal procedure to extract
the greatest amount of work from a system operating under well-defined constraints. For externally controlled small systems, the optimal driving
protocol maximizes the mean work spend in a finite-time transition between two given system states. For simplicity we consider an externally
controlled single level system, which is embedded in a thermal environment and coupled to a particle reservoir. The optimal protocols are
calculated from a master equation approach for different system-reservoir couplings. For open systems, the system-reservoir couplings are
shown to have a striking influence on the optimal driving setup. We point out that the optimal protocols have discontinuous jumps at the
initial and final times. Finally, this work provides a first attempt to extend these calculations to larger system sizes.
\end{abstract}

\section{Introduction}
\label{sec:intro}
A good theoretical understanding of the optimal control of energy conversion processes is a prerequisite for tailoring
efficient artificial nanodevices for specific needs. Typical examples are soft and biomatter systems,
such as Brownian or molecular motors, organic photovoltaic solar cells and quantum dots.
With minimization of the system size thermal fluctuations become relevant and the non-equilibrium
behavior of such systems depends strongly on the driving forces
and the changes of one system state to another, which are inherently finite in time.
Thus, thermodynamic processes take place in finite time and the thermodynamic quantities like heat and work are
now random but still fulfill a stochastic energy balance. In these systems it is useful to
introduce microscopic heat and work quantities as random variables whose averages lead to
the common thermodynamic quantities. Averages over functions of these microscopic
heat and work quantities yield generalized fluctuation theorems (for reviews, see \cite{Seifert:2012}, and references therein).
A common feature of many artificial nano-sized devices, where fluctuation theorems can be applied, is that those are
mostly driven by time-varying external fields (often called as protocol) or electrochemical potential differences.

Of particular importance of this class of non-equilibrium systems is to identify the optimal procedure to
extract the greatest amount of work from the device operating under given constraints \cite{Schmiedl/Seifert:2007}.
Only a few studies so far have addressed the problem of identifying the optimal protocol that yields the minimum amount
of work done on the system, which is required to drive a nano-scale system from one equilibrium state to another
in finite time \cite{Schmiedl/Seifert:2007,Then/Engel:2008,Esposito/etal:2010,Esposito/etal:2010b}. Note that this formulation is
in reverse to the quest for the optimal protocol that provides the greatest amount of work from the artificial device.
Whether on a continuous (Langevin equation) or a discrete (master equation) state space, a surprising result of all these studies
is that the optimal protocol exhibits sudden jumps at the beginning and at the end of the
thermodynamic process, while in between the optimal protocol varies smoothly. It can be argued that the initial jump
in the optimal protocol is an immediate jump from equilibrium to a stationary state and the final jump
allows a slower driving of the system at earlier times.

Here, we address a similar question on optimal driving setups. The focus of this work is
on the couplings of the system to reservoirs, which have a strong influence on the specific form of the optimal protocol.
To keep the analysis conceptual simple, we consider as working medium a two-state system driven by a time-dependent protocol,
which is often called as one of the prototype models in non-equilibrium statistical mechanics. We apply an optimization procedure
based on a variational analysis. This allows us to get an analytic expression of the optimal protocol with respect to
different system-reservoir couplings.

\section{Model}
\label{sec:model}
We consider a single level system as working medium of an artificial nanodevice, in
contact with a particle reservoir at temperature $T$, which is characterized by the chemical potential $\mu$.
In general, the temperature $T$ and the chemical potential $\mu$ may also be time dependent.
The site energy $\varepsilon(t)$ is assumed to be time-dependent, which can be modified between an initial value $\varepsilon_0$
and a final value $\varepsilon_1$ by an external agent according to a given protocol. In what follows, it is convenient to introduce the energy difference
$\epsilon(t) = \varepsilon(t) - \mu (t)$. The population of the single level at time $t$ is characterized by the occupation probability $p(t)$.
The system dynamics is modeled by a master equation approach with time-dependent rates $w_1 (t)$ and $w_2(t)$
\cite{Chvosta/etal:2010,Einax/etal:2009,Einax/etal:2010a,Einax/etal:2010b,Einax/etal:2011} accounting for the time evolution of the occupation probability $p(t)$,
\begin{align}
\label{eq:master_equ}
\dot{p} = - w_1 (t) p(t) - w_2 (t) (1-p(t))\, .
\end{align}
We assume that these transition rates obey detailed balance at each time instant.
Due to the time-varying of the site energy $\epsilon(t)$, positive or negative energy flows into the system. From the thermodynamic point of view
the time derivative of the internal energy $E(t)=\epsilon(t)p(t)$ of the artificial nanodevice is the sum of a work flow
$\dot{W} = \dot{\epsilon} (t) p(t)$ and a heat flow $\dot{Q} = \epsilon(t) \dot{p} (t)$.
Consequently, during the process time $\tau$ the energy change in the system obeys the statistical mechanics formulation of the first law of thermodynamics,
$\Delta E(\tau) = Q(\tau) + W(\tau)$, where $\Delta E (\tau) = \int_{0}^{\tau} dt\, \dot{E} (t) = E(\tau)-E(0) \equiv \epsilon (\tau) p(\tau) - \epsilon (0) p(0)$
with $\epsilon (0)=\epsilon_0$ and $\epsilon (\tau)=\epsilon_1$. The occupation probability of the equilibrium state at the beginning is
$p(0)=[\exp(\beta \epsilon_0) +1]^{-1}$ with $\beta = 1/k_{\rm B} T$. Accordingly, we have $W=\int_{0}^{\tau} dt\,\dot{\epsilon} (t) p(t)$ and
$Q= \int_{0}^{\tau} dt\,\epsilon(t) \dot{p}(t)$. Both work $W$ and heat $Q$ can be interpreted as functionals of the occupation
probability and thus depend, in particular, on the realized transition path. If $W(t) < 0$, the (positive) work $-W(t)$ is done by the system on the environment.

The details of the system-reservoir coupling determine decisive the exchange of particles between the reservoir and the system~\cite{Dierl/etal:2012}. This is
reflected in the form of the transition rates. In order to proceed, we need to specify these rates for configurational transitions
consistent with the detailed balance condition. For example, in a quantum dot coupled to a metal lead,
the rates $w_1 (t)$ and $w_2(t)$ are Glauber (Fermi) rates, i.~e., $w_1 (t)=C [\exp(-\beta \epsilon(t))+1]^{-1}$ and
$w_2 (t)= C [\exp(+\beta \epsilon(t))+1]^{-1}$, respectively \cite{Esposito/etal:2010}. $C$ is the inverse of a characteristic time scale involve in the exchange
of particles between the reservoir and the system.
We adopt here symmetric rates $w_1(t)=C e^{\beta \epsilon(t)/2}$ and $w_2(t)= C e^{-\beta \epsilon(t)/2}$.
Consequently, we have $w_1(t) + w_2(t) = 2 C \cosh [\beta \epsilon(t)/2]$.
Symmetric transition rates are widely used in biomatter or ionic systems \cite{Fisher/Kolomeisky:1999,Einax/etal:2010a}.

\section{Calculation of optimal driving protocols}
\label{sec:calculation}
In order to calculate the optimal driving protocol $\epsilon(t)$ which minimizes the work for the given constraints
$\epsilon_0$, $\epsilon_1$, $p(0)$, and $\tau$, we apply the proposed procedure given in Reference~\cite{Esposito/etal:2010}.
To minimize $W=\Delta E - Q$ we have to minimize $\Delta E$ and maximize $Q$ simultaneously
Since $\Delta E$ depends only on $p(\tau)$, we need to maximize the heat $Q$. The essential steps of this procedure are: {\it (i)}
find the protocol that gives the maximum heat $Q$ for a given value of $p(\tau)$ and {\it (ii)} conduct the optimization with respect to the
final state $p(\tau)$. To this end we rewrite the heat as
$Q = \int_0^\tau dt\, L(p,\dot{p})$ with $L(p,\dot{p})=\epsilon (t) \dot{p}$, because
$\epsilon(t)$ can be expressed as function of $p(t)$ and $\dot{p}(t)$. A variational analysis $\delta \int L dt =0$ leads to the
Euler-Lagrange equation $L - \dot{p} \frac{\partial L }{\partial \dot{p}} \equiv - \dot{p}^2 \frac{\partial \epsilon}{\partial \dot{p}} = \tilde{K}$.
Here $\tilde{K}$ is the constant of integration. For symmetric rates the Euler-Lagrange equation has the form
\begin{align}
\label{eq:p_dot_nonequ}
\pm \frac{2 (\dot{p}/C)^2}{\sqrt{ (\dot{p}/C)^2 +4p(1-p)}} &= {K} \, ,
\end{align}
where ${K} =\tilde{K}\beta/C$. Equation (\ref{eq:p_dot_nonequ}) is a quartic equation with respect to $\dot{p}$, which has two
real (physical) solutions $\dot{p}/C = \pm \frac{1}{4} \sqrt{ 2 K^2 + 2 \sqrt{K^4 + 64 K^2 p(1-p)}}$ and two imaginary solutions.
From these two first-order differential equations one can calculate $p(\tau)$ for given $p(0)$ and $\tau$ via
\begin{align}
\label{eq:p_tau}
C \tau &= \pm \int_{p(0)}^{p(\tau)} \frac{dp}{\frac{1}{4} \sqrt{ 2 K^2 + 2 \sqrt{K^4 + 64 K^2 p(1-p)}}}
\end{align}
as function of $K$. Subsequently, $p(\tau)|_K$ can be used to maximize the heat $Q$ with respect to $K$ by performing the integration
\begin{align}
\label{eq:heat_max}
\beta Q|_K &= \beta \int_0^{\tau} dt\,\epsilon(t) \dot{p}(t) \equiv \int_{p(0)}^{p(\tau)} \beta \epsilon(p) dp\, .
\end{align}
Here, the $K$-dependent protocol $\epsilon$ can be rewritten with respect to $p$. For this purpose, we start from the quartic equation
$4 (\dot{p}/C)^4 = K^2 [(\dot{p}/C)^2 +4p(1-p)]$. Using the evolution equation~(\ref{eq:master_equ}) with symmetric rates and multiplying
the quartic equation with $\exp{(2\beta\epsilon)}$, the substitution $t^{m}=\exp(m\beta \epsilon)$; $m=1,2,3,4$ then gives
the polynomial equation of 4th order $t^4 + c_3t^3 + c_2t^2 + c_1 t + c_0 =0$. The coefficients are
$c_3= \left(16p^2-16p-K^2\right)/(4p^2)$, $c_2= \left(p-1\right)\left(12p^2-12p+K^2\right)/(2p^3)$,
$c_1= \left(p-1\right)^2\left(16p^2-16p-K^2\right)/(4p^4)$, and $c_0= \left(p-1\right)^4/p^4$.
Since the relation $(c_1/c_3)^2=c_0$ holds between the coefficients $c_0$, $c_1$, and $c_3$, this quartic equation can be rewritten
in a quasi-symmetric equation $\left(t^2+\frac{c_0}{t^2}\right) + \frac{c_1}{\sqrt{c_0}} \left( t+\frac{\sqrt{c_0}}{t} \right) +c_2 =0$.
Using the transformation $z=t+\sqrt{c_0}/t$, this quasi-symmetric equation can be reduced to two quadratic equations.
This leads to two real roots of the form
\begin{align}
\label{eq:E_p}
\beta \epsilon (p) &= \ln \left[ \frac{ A(p,K)+B(p,K)}{16 p^2} \pm \frac{ \sqrt{2K^2 D(p,K)+2 A(p,K) B(p,K)} }{16 p^2} \right] \, ,
\end{align}
where $A(p,K) = K^2+16p(1-p)$, $B(p,K) = \sqrt{K^4+64K^2p(1-p)}$, and $D(p,K) = K^2+48p(1-p)$. In addition, we have two nonphysical complex conjugate roots.
Finally, we are able to optimize the work $W|_K=\Delta E|_K - Q|_K=\epsilon_1 p(\tau)|_K - \epsilon_0 p(0) - Q|_K$
with respect to $K$.

 \section{Results and discussion}
\label{sec:results}
\begin{figure}[h]
 \centering
 \includegraphics[width=0.4\columnwidth]{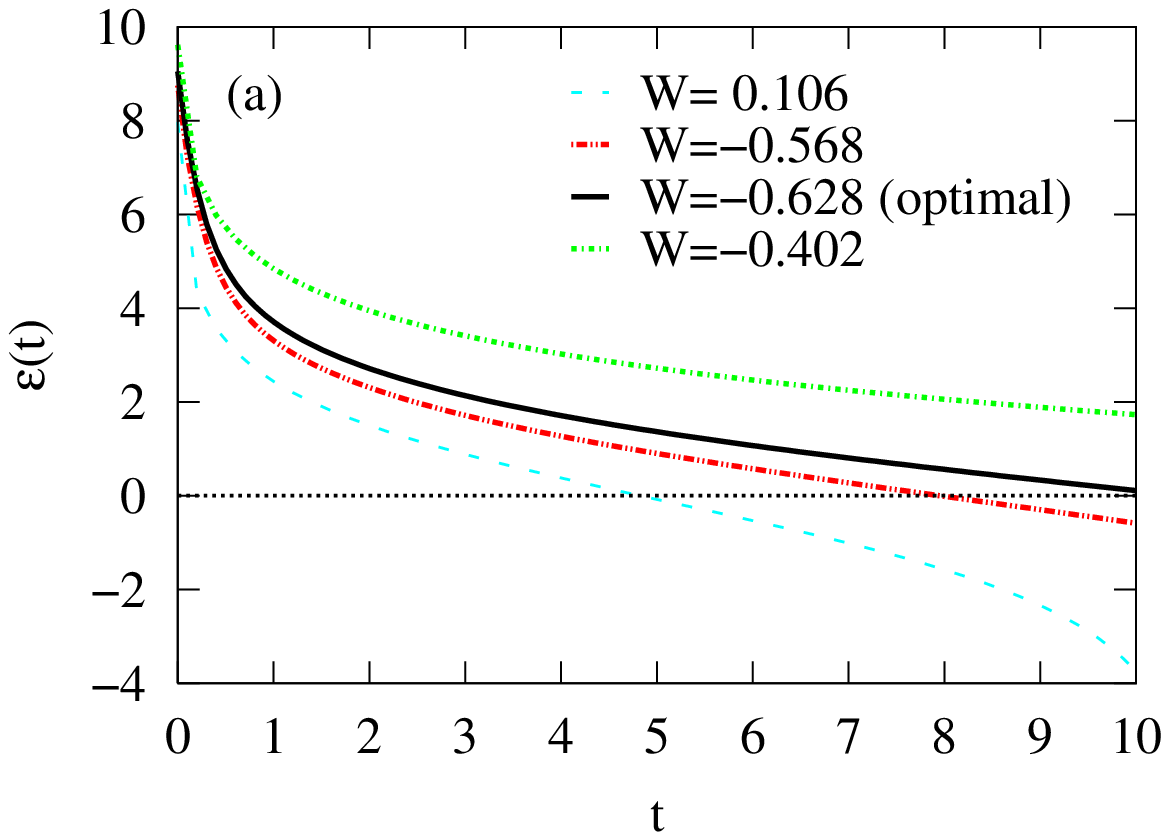}
 \includegraphics[width=0.4\columnwidth]{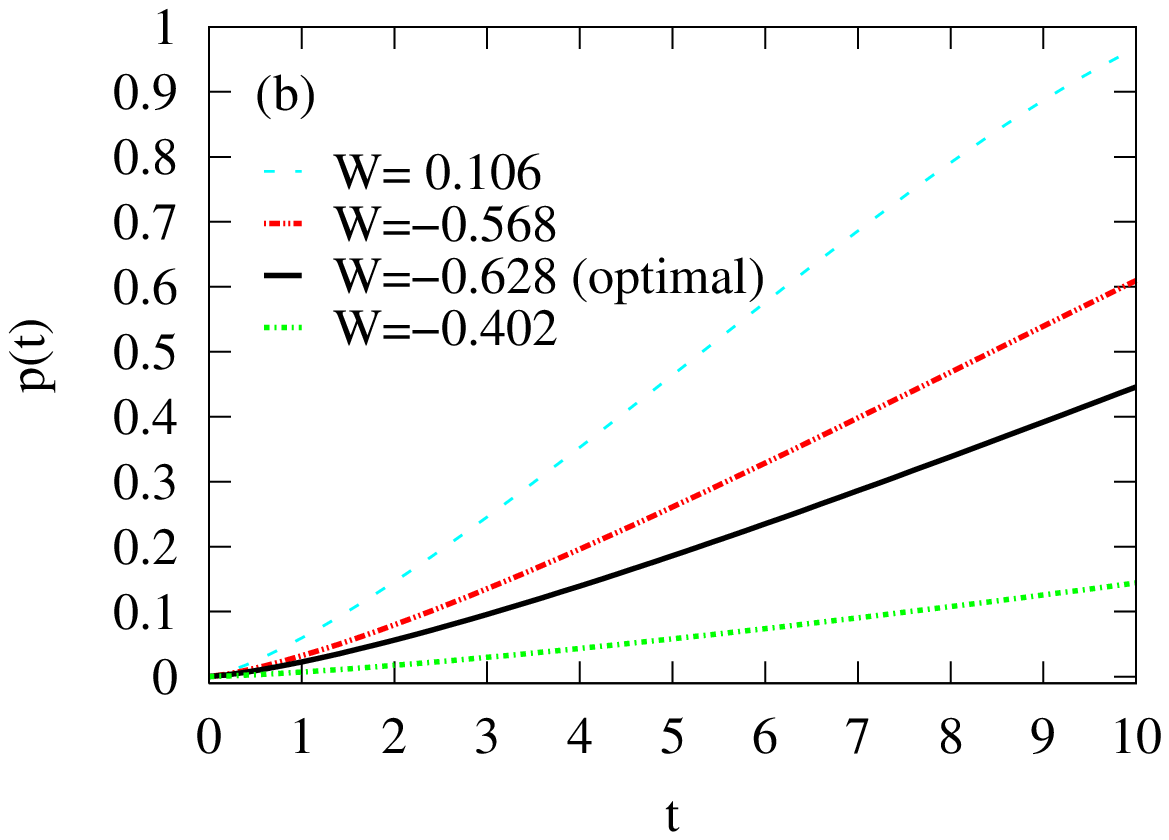}\\
 \includegraphics[width=0.4\columnwidth]{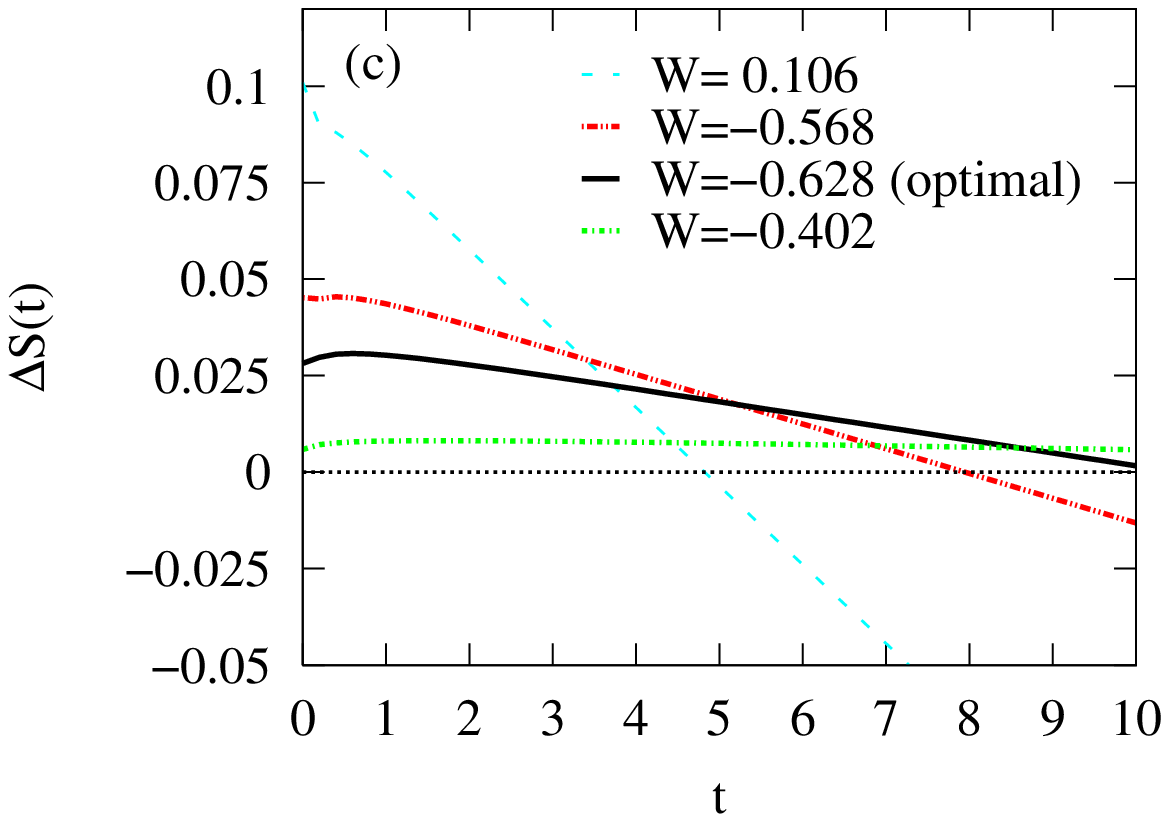}
 \includegraphics[width=0.4\columnwidth]{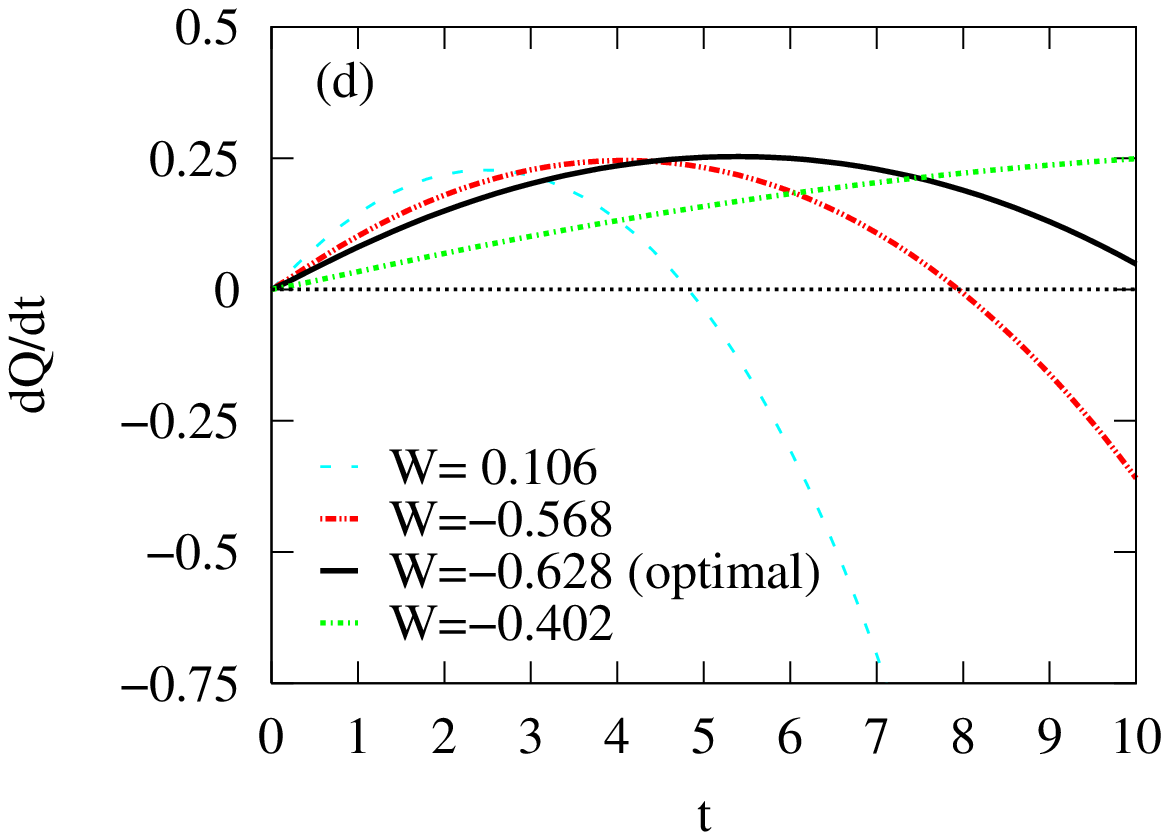}
 \caption{System properties for the case, in which the energy level is lowered from $\epsilon_0=10$ to $\epsilon_1=0$ as function of time $t$
 starting from $t_0=0$ until $t_1=10$; (a) protocol and (b) occupation probability. An optimal protocol (black solid line) can be identified
 with minimum cumulated work $W=-0.628$ for $K_{\rm opt}=0.00591$ and the total available time $\tau=t_1-t_0=10$.}
 \label{fig:fig1}
\end{figure}
In the calculation discussed below, the following choice of parameters was used: $C=1$, $\beta=1$, $\epsilon_0=10$, and $\epsilon_1=0$.
Let us first take a closer look at the optimal protocol for symmetric rates as function of the processing time $t$ for different values of $K$.
To this end we calculate Equation~(\ref{eq:E_p}) and Equation~(\ref{eq:heat_max}) with the upper sign (+) by using Equation~(\ref{eq:p_tau}) with the lower sign (-) for downward processes. 
In this case, work is done by the system on the environment.
\begin{wrapfigure}{r}{0.45\textwidth}
\hspace*{1ex}
\includegraphics[width=0.45\textwidth]{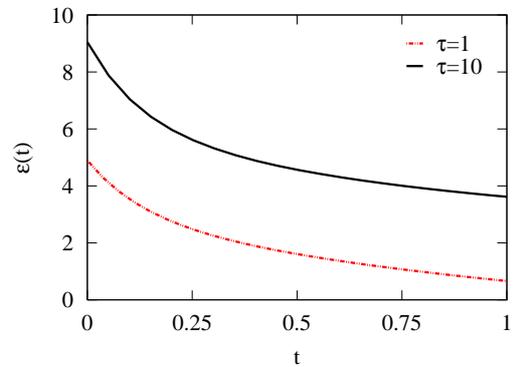}
\caption{The optimal protocol for $\tau=1$ ($K_{\rm opt}=0.17$) and $\tau=10$ ($K_{\rm opt}=0.00591$) as function of time $t$.}
\label{fig:fig2}
\end{wrapfigure}
Figure~\ref{fig:fig1}(a) shows both the optimal protocol (black solid line) and three non-optimal protocols as function of the processing time $t$.
We can clearly identify jumps of the protocol at the beginning and at the end of the process. As seen in Figure~\ref{fig:fig1}(b), an initially
almost empty systems is populated with particles with increasing time. The protocol controls how fast this single level can be occupied.
In addition, Figure~\ref{fig:fig1}(c) and (d) show other characteristic quantities of the system, such as the entropy change $\Delta S = \beta Q$
and the heat flow $\dot{Q}$, respectively.
Figure~\ref{fig:fig2} shows the optimal protocol for fast ($\tau=1$) and slow ($\tau=10$) processes. We observe that the associated value of
$K_{\rm opt}$ is larger for fast processes than for slow processes.

Thus, to understand the meaning of the parameter $K$, we square equation~(\ref{eq:p_dot_nonequ}). By using evolution equation (\ref{eq:master_equ})
with symmetric rates we arrive, after sorting with respect to $p$, at a quartic equation $a_4p^4 + a_3p^3 + a_2p^2 + a_1 p + a_0 =0$.
The coefficients are $a_4=4e^{2\beta \epsilon} + 4e^{-2\beta \epsilon} + 16e^{\beta \epsilon} + 16e^{-\beta \epsilon} + 24$,
$a_3=-16e^{-2\beta \epsilon} - 16e^{\beta \epsilon} - 48e^{-\beta \epsilon} - 48$,
$a_2=24e^{-2\beta \epsilon}+ 48e^{-\beta \epsilon}+ 24 + K^2 (2-e^{\beta \epsilon}-e^{-\beta \epsilon})$,
$a_1=-16e^{-2\beta \epsilon} - 16e^{-\beta \epsilon} + K^2 (-2+2e^{-\beta \epsilon})$, and
$a_0=4e^{-2\beta \epsilon}-K^2 e^{-\beta \epsilon}$.
The roots of this fourth order polynomial equation can be obtained analytically. To this end, we perform
the transformation $p=u-a_3/a_4$. The ratio
$-a_3/a_4=[\exp (\beta \epsilon(t))+1]^{-1}$
is even the thermal equilibrium distribution. This substitution eliminates the $p^3$ term
in 4th order polynomial equation leading to the depressed quartic equation
$u^4 - b_2u^2 - b_1 u + b_0 =0$. Here,
\begin{align*}
b_2=  \frac{K^2}{8} \frac{\tanh^2 (\beta \epsilon/2)}{1+\cosh (\beta \epsilon)}\, ,
b_1=  \frac{K^2}{4} \frac{\tanh (\beta \epsilon)/2)}{(1+\cosh (\beta \epsilon)))^2}\, ,
b_0= -\frac{K^2}{8} \frac{1}{(1+\cosh (\beta \epsilon)))^3} \, .
\end{align*}
Realizing that $b_0=-b_1^2/(4b_2)$, the depressed quartic equation can be written in the form
$u^4 = \left( \sqrt{b_2} u + \frac{b_1}{2 \sqrt{b_2}}\right)^2$, which has two real and two complex conjugated roots.
The two real roots $u_{\pm} (\epsilon,K)$ are determined by the branch $u^2 - \sqrt{b_2} u - \frac{b_1}{2 \sqrt{b_2}}=0$ leading to
te expression
\begin{align}
\label{eq:p_non_equ}
p(t) &= \frac{1}{e^{\beta \epsilon}+1} \left( 1+ \frac{K}{4} e^{\beta \epsilon/2} \tanh (\beta \epsilon/2)
\pm \sqrt{\frac{K^2}{8} e^{\beta \epsilon} \tanh^2 (\beta \epsilon/2)+ \frac{K}{2} \frac{e^{\beta \epsilon}}{\cosh (\beta \epsilon/2)} } \right) \, .
\end{align}
Equation~(\ref{eq:p_non_equ}) expresses the occupation probability as function of the protocol $\epsilon$ and $K$.
If we assume a different system-reservoir coupling, say Fermi rates, the form of $p(t)$ \cite{Esposito/etal:2010} is quit different from
those given in Equation~(\ref{eq:p_non_equ}). This emphasizes that the nature of the system-reservoir coupling has a striking influence on the optimal driving setup.
If $K=0$, $p(t)$ is the equilibrium distribution of a single level system. This implies that $K=0$ coincides with a
quasistatic process control, i.~e.,  to drive a artificial nanodevice from one equilibrium state to another equilibrium state.
Consequently, the parameter $K$ quantifies how far the system differs from the quasistatic limit.

In order to extent these investigations to large systems, we consider the simplest case, in which a system with $N$ sites and energies $\epsilon_l(t)$ interacts with a particle reservoir at temperature $T$. Particles can only enter or leave each level and we assume that no transitions between site $l$ and site $k$ take place. Thus, the internal energy of the system is given by an additive principle, $E(t) = \sum_l \epsilon_l(t) p_l(t)$, and the time-derivative of $E(t)$ yields $\dot{W} + \dot{Q}$ with $\dot{W}=\sum_l \dot{W}_l = \sum_l \dot{\epsilon}_l (t) p_l(t)$ and
$\dot{Q} = \sum_l \dot{Q_l} = \sum_l \epsilon_l (t) \dot{p}(t)$. In this special case the application of the variational analysis presented
in Section~\ref{sec:calculation} is straightforward, in order to get the optimal setup of large systems. This fails if we allow
an exchange of particle between site $l$ and site $k$.


\section*{References}
\begin{thebibliography}{12} 
\bibitem{Seifert:2012} Seifert U 2012 {\it Rep. Prog. Phys.} {\bf 75} 126001
\bibitem{Schmiedl/Seifert:2007} Schmiedl T and Seifert U 2007 {\it Phys. Rev. Lett.} {\bf 98} 108301
\bibitem{Then/Engel:2008} Then H and Engel A 2008 {\it Phys. Rev. E} {\bf 77} 041105
\bibitem{Esposito/etal:2010} Esposito M, Kawai R, Lindenberg K and Van den Broeck C 2010 {\it Europhys. Lett.} {\bf 89} 20003
\bibitem{Esposito/etal:2010b} Esposito M, Kawai R, Lindenberg K and Van den Broeck C 2010 {\it Phys. Rev. E} {\bf 81} 041106
\bibitem{Chvosta/etal:2010} Chvosta P, Einax M, Holubec V, Ryabov A and Maass P, 2010 {\it J. Stat. Mech.} P03002
\bibitem{Einax/etal:2009} Einax M and Maass P 2009 {\it Phys. Rev. E} {\bf 80}, 020101
\bibitem{Einax/etal:2010a} Einax M, K\"orner M, Maass P and Nitzan A 2010 {\it Phys. Chem. Chem. Phys.} {\bf 12} 645
\bibitem{Einax/etal:2010b} Einax M, Solomon G C, Dieterich W and Nitzan A 2010 {\it J. Chem. Phys.} {\bf 133} 054102
\bibitem{Einax/etal:2011} Einax M, Dierl M and Nitzan A 2011 {\it J. Phys. Chem. C} {\bf 115} 21396
\bibitem{Dierl/etal:2012} Dierl M, Maass P and Einax M 2012 {\it Phys. Rev. Lett.} {\bf 108} 060603
\bibitem{Fisher/Kolomeisky:1999} Fisher M E and Kolomeisky A B 1999 {\it Proc. Natl. Acad. Sci.} {\bf 96} 6597
\end{thebibliography}
\end{document}